\documentclass[manuscript]{aastex}
\usepackage{graphicx}
\usepackage{epstopdf}
\usepackage{natbib}
\usepackage[lofdepth,lotdepth,caption=false]{subfig}
\usepackage{hyperref}
\usepackage{float}

\begin{document}

\title{Lingering grains of truth around comet 17P/Holmes}

\author{R. Stevenson*\altaffilmark{,1}, J. M. Bauer\altaffilmark{1,2}, E. A. Kramer\altaffilmark{3}, T. Grav\altaffilmark{4}, A. K. Mainzer\altaffilmark{1}, J. R. Masiero\altaffilmark{1}}

\altaffiltext{*}{To whom correspondence should be address: Rachel.A.Stevenson@jpl.nasa.gov}
\altaffiltext{1}{Jet Propulsion Laboratory, California Institute of Technology, 4800 Oak Grove Drive, Pasadena, CA 91109}
\altaffiltext{2}{Infrared Processing and Analysis Center, California Institute of Technology, Pasadena, CA 91125}
\altaffiltext{3}{Dept. of Physics, University of Central Florida, 4000 Central Florida Blvd, Orlando, FL 32816}
\altaffiltext{4}{Planetary Science Institute, 1700 East Fort Lowell, Suite 106, Tucson, AZ 85719-2395}

\begin{abstract}

Comet 17P/Holmes underwent a massive outburst in 2007 Oct., brightening by a factor of almost a million in under 48 hours. We used infrared images taken by the Wide-Field Survey Explorer mission to characterize the comet as it appeared at a heliocentric distance of 5.1~AU almost 3 years after the outburst. The comet appeared to be active with a coma and dust trail along the orbital plane. We constrained the diameter, albedo, and beaming parameter of the nucleus to 4.135 $\pm$ 0.610 km, 0.03 $\pm$ 0.01 and  1.03 $\pm$ 0.21, respectively. The properties of the nucleus are consistent with those of other Jupiter Family comets. The best-fit temperature of the coma was 134~$\pm$~11~K, slightly higher than the blackbody temperature at that heliocentric distance. Using Finson-Probstein modeling we found that the morphology of the trail was consistent with ejection during the 2007 outburst and was made up of dust grains between 250~$\mu$m and a few cm in radius. The trail mass was $\sim$ 1.2~-~5.3~$\times$~10$^{10}$~kg. 

\end{abstract}

\section{Introduction}

Comet 17P/Holmes (hereafter 17P) has undergone 3 massive outbursts since its discovery in 1892 \citep{1892Obs....15..441H}, most recently brightening by a factor of almost a million and becoming visible to the naked eye in 2007 Oct.  The outbursts are likely thermally-driven since all three occurred 6-9 months after 17P passed through perihelion.  Dynamically, 17P appears to be a typical Jupiter family comet (JFC) with a semi-major axis of 3.62 AU, eccentricity of 0.43, and inclination of 19$^{\circ}$.1.  The comet is enigmatic given its propensity for unusually large outbursts but dynamical and physical properties similar to other JFCs.

The material ejected during the 2007 outburst included gas species, dust grains, and macroscopic fragments (e.g. \citealt{2008ApJ...680..793D,2009P&SS...57.1162C,2009AJ....137.4538Y,2010Icar..208..276R,2010AJ....139.2230S}). Much of the smaller dust expanded in an almost spherical shell around the nucleus, while larger dust grains were observed to separate as a ``blob'' at a slower rate of $\sim$~120~-~135~m~s$^{-1}$ (e.g. \citealt{2008A&A...479L..45M,2009AJ....138..625L,2010MNRAS.407.1784H}). \cite{2010Icar..208..276R} and \cite{2012A&A...542A..73B}  detected a slower moving core-component of the largest grains that separated from the nucleus at a relative velocity of $\sim$ 7-9 m s$^{-1}$. Large dust grains may persist in the vicinity or along the trail of a comet for years after ejection from the nucleus \citep{1990Icar...86..236S,1998ApJ...496..971L,2011ApJ...738..171B}. In this work we used infrared (IR) images obtained with the Wide-Field Infrared Survey Explorer (WISE) to examine the evolution of 17P several years after the 2007 outburst.

\section{WISE Observations and Reduction}
\label{sec:wobs}

The WISE telescope launched in 2009 Dec.\ and conducted an all-sky survey over the following year.  The 40 cm telescope covered a 47$^{\prime}$ $\times$ 47$^{\prime}$ field of view (FOV) in four IR bands simultaneously.  The bands had central wavelengths of 3.4~$\mu$m, 4.6~$\mu$m, 12~$\mu$m, and 22~$\mu$m, and are referred to as W1, W2, W3, and W4, respectively.  The median pixel scale in bands W1, W2, and W3 was 2$^{\prime\prime}$.8 pixel$^{-1}$, while 2~$\times$~2 binned W4 images had a pixel scale of 5$^{\prime\prime}$.5 pixel$^{-1}$ \citep{2010AJ....140.1868W}.  The effective exposure times were 7.7~s for W1 and W2 images, and 8.8~s for W3 and W4 images.   The individual exposures and extracted sources from each frame were archived and searched using tools developed as part of the NEOWISE project \citep{2011ApJ...736..100M}.

The data were initially processed by the WISE Science Data System, which removed the instrumental signatures and provided astrometric and photometric calibration.  Astrometric accuracy was $\sim$ 0$^{\prime\prime}$.2, while absolute photometric accuracy was $\sim$~5-10\% \citep{2011ApJ...736..100M,2012wise.rept....1C}.  

WISE pointed towards 17P a total of 14 times during the mission. Three of these sets of images were obtained after the cryogen had been depleted and thus only produced images at the two shortest wavelengths. Here we use the 11 sets of images in all 4 bands that were 
obtained within 24 hours between UT 2010 May 14 and 15 (Table~\ref{table:obs}).  At this time 17P was at a heliocentric distance of 5.1 AU and a true anomaly of $\sim$ 170$^{\circ}$, approximately 5 months prior to reaching aphelion.  These frames were aligned using the predicted orbital motion of the comet as calculated by the JPL Horizons ephemeris service and combined using the AWAIC (A WISE Astronomical Image Co-Adder) stacking algorithm \citep{2009ASPC..411...67M}, which includes an outlier-rejection algorithm.  By using the stacked images for this work, we increased the signal to noise ratio, and average over rotational variations, which may amount to 0.3 mag in R-band \citep{2006MNRAS.373.1590S}.
  The stacked images were resampled to have pixel scales of 1$^{\prime\prime}$ pixel$^{-1}$, corresponding to a projected on-sky distance of 3600 km pixel$^{-1}$, with PSFs having average full-width half-maxima (FWHM) of 6$^{\prime\prime}$.1, 6$^{\prime\prime}$.4, 6$^{\prime\prime}$.5, and 12$^{\prime\prime}$.0 in bands W1, W2, W3, and W4, respectively \citep{2010AJ....140.1868W}.  Though the spacecraft did not track the comet's motion, trailing is not a concern since the maximum motion of the comet during an exposure was 0.03$^{\prime\prime}$, significantly less than the FWHM or pixel scale of any image.

\begin{deluxetable}{cccccc}
\tabletypesize{\scriptsize}
\tablecaption{Observations}
\tablewidth{0pt}
\tablehead{
\colhead{ } & \colhead{Date [UT]} & \colhead{Number of Frames} &  \colhead{r$_{H}$ [AU]} & \colhead{$\Delta$ [AU]} & \colhead{$\alpha$ [deg]}  }
\startdata
17P/Holmes & 2010 May 14-15 & 11 & 5.13 & 4.93 & 11.3 \\
\enddata
\label{table:obs}
\end{deluxetable}

To convert counts to fluxes we used the instrumental zero points given in \cite{2010AJ....140.1868W}. We revised the zero points by -8\% in W3 and +4\% in W4 to account for the observed discrepancy between red and blue calibrators.  We corrected for the loss of light outside of fixed apertures by using the aperture corrections given in \cite{2011wise.rept....1C}.  These amounted to -0.34 and -0.65 mag for W3 and W4 for apertures of radius 11$^{\prime\prime}$.   In this work we chose to use apertures with radii of 11$^{\prime\prime}$ as a compromise between limiting the intrusion of background signal into the aperture and still capturing the majority of the PSF.  Finally, we performed a color correction, which was necessary due to the wide band pass of the filters. We calculated the correction by interpolating the color corrections given for a range of temperatures in \cite{2010AJ....140.1868W} to the estimated blackbody temperature of 17P. 

\begin{figure}[h]
\centering
\subfloat{\includegraphics[totalheight=7cm]{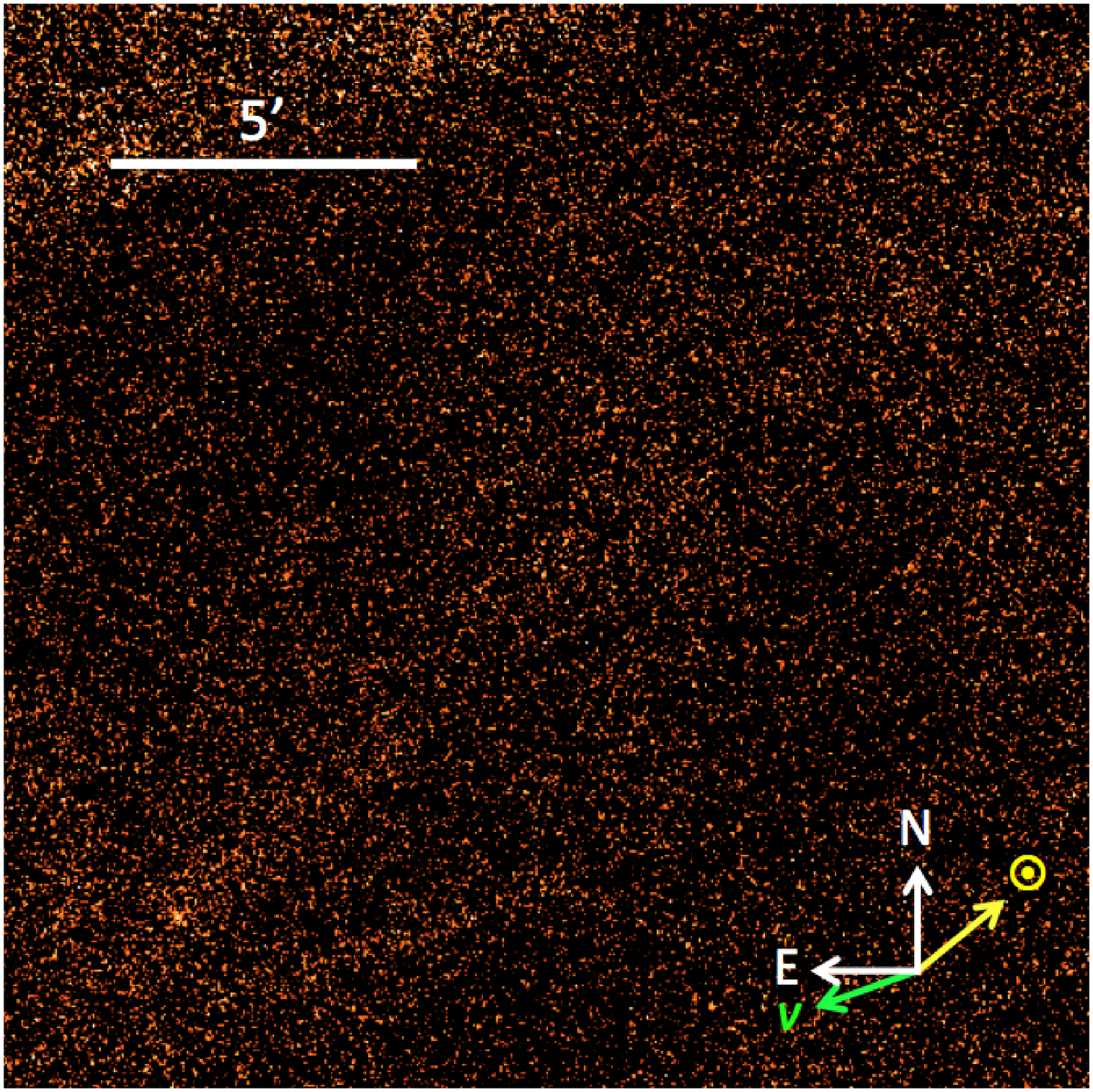}}
\hspace{2bp}
\subfloat{\includegraphics[totalheight=7cm]{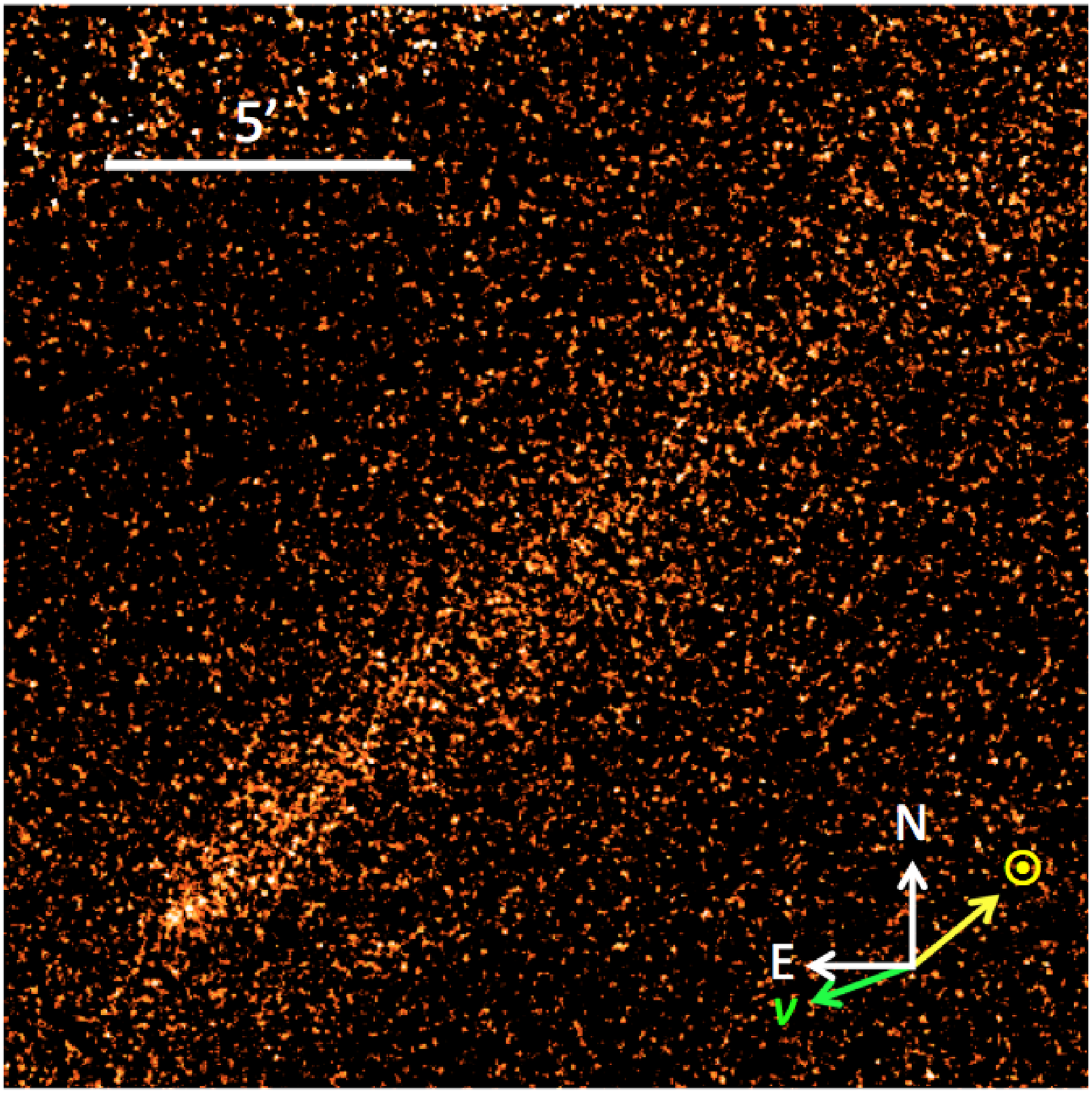}}
\caption{Comet 17P/Holmes as observed by the WISE mission in W3 (left; 12~$\mu$m) and W4 (right; 22~$\mu$m) on UT 2010 May 14-15.  The nucleus is located in the south-east corner of the image. Celestial north ($N$) and east ($E$) are marked, as are the solar ($\odot$) and velocity ($v$) vectors.}
\label{fig:holmes}
\end{figure}

\section{Results}

17P was detected in the longer wavelength W3 and W4 bands (Figure~\ref{fig:holmes}) but was not detected in bands W1 or W2.  A dust trail was seen in both W3 and W4 images, though it was considerably brighter in W4.   The modal average of the background was calculated $\sim$~5$^{\prime}$ from the nucleus over an area of $\sim$~2.2 square arcminutes and subtracted from the stacked images. We set 5~$\sigma$ upper limits on the signal from 17P in the W1 and W2 bands of 0.03~mJy and 0.10~mJy, respectively, using an 11$^{\prime\prime}$ radius aperture.  The total fluxes within 11$^{\prime\prime}$ radius apertures were 0.77~$\pm$~0.15~mJy and 8.97~$\pm$~1.91~mJy in bands W3 and W4, respectively.  Table~\ref{table:fluxes} shows the fluxes measured for 17P and the fluxes obtained by best-fit thermal models to the nucleus and coma signals (discussed in sections~\ref{sec:nuc} and \ref{sec:coma}, respectively).

\begin{deluxetable}{ccccc}
\tabletypesize{\scriptsize}
\tablecaption{Fluxes in mJy}
\tablewidth{0pt}
\tablehead{
\colhead{} & \colhead{W1} & \colhead{W2} & \colhead{W3} & \colhead{W4}\\
\colhead{} & \colhead{(3.4 $\mu$m)} & \colhead{(4.6 $\mu$m)} & \colhead{(12 $\mu$m)} & \colhead{(22 $\mu$m)}}
\startdata
17P/Holmes, total flux & $<$ 0.03 & $<$ 0.10 & 0.77 $\pm$ 0.15 & 8.97 $\pm$ 1.91\\ 
Nucleus (measured) & $<$ 0.03 & $<$ 0.10 & 0.33 $\pm$ 0.05 & 1.56 $\pm$ 0.31\\
Nucleus (best-fit) & 1.92 $\times$ 10$^{-4}$ & 2.04 $\times$ 10$^{-4}$ & 0.27 & 1.68\\
Coma (measured) & $<$ 0.03 & $<$ 0.10 & 0.45 $\pm$ 0.09 & 7.41 $\pm$ 1.58\\
Coma (best-fit) & 5.91 $\times$ 10$^{-3}$ & 3.59 $\times$ 10$^{-3}$ & 0.45 & 7.41\\
\enddata
\label{table:fluxes}
\end{deluxetable}

\clearpage

\subsection{Nucleus}
\label{sec:nuc}

\begin{figure}[h]
\centering
\subfloat{\includegraphics[totalheight=8cm]{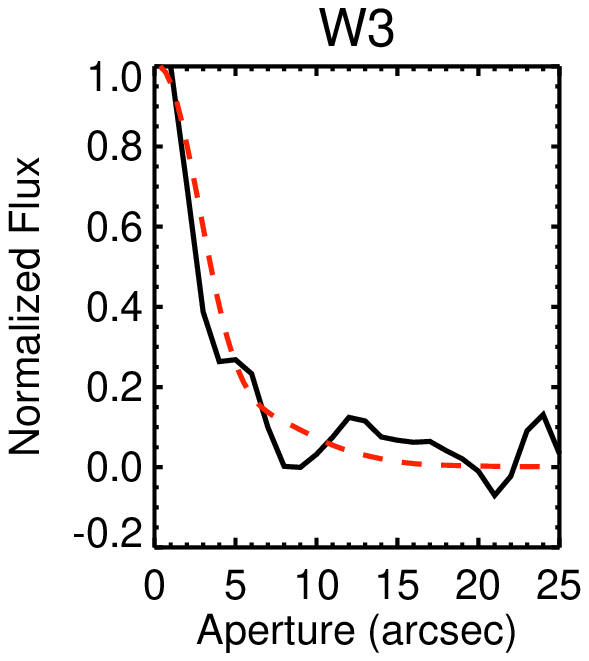}}
\hspace{2bp}
\subfloat{\includegraphics[totalheight=8cm]{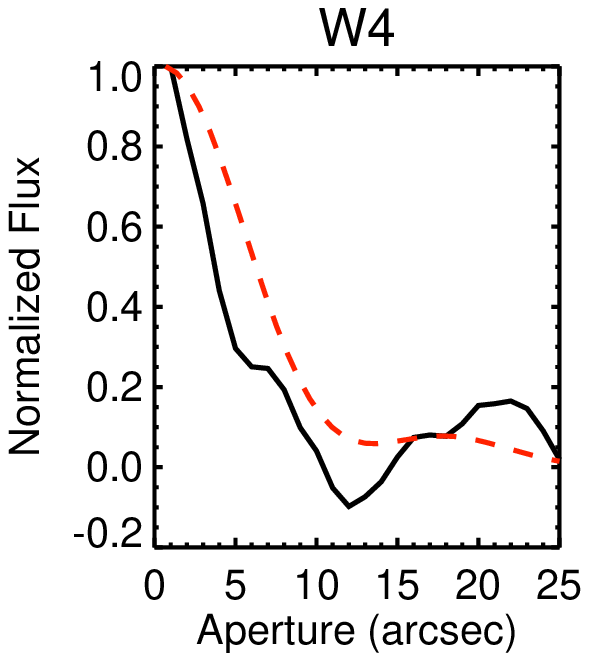}}
\caption{Surface brightness profile of the signal that remained following coma subtraction (black solid line) and the WISE PSF (red dashed line) for bands W3 (left) and W4 (right).}
\label{fig:PSFcomp}
\end{figure}

The total signal was a mix of contributions from the nucleus and dust particles in the coma and trail around the nucleus.  We separated those signals by fitting the non-PSF-like signal with an analytical function of the form F $\rho^{-n}$, where F is a scalar, $\rho$ is the distance from the nucleus, and $n$ is a power law index \citep{1999PhDT.........8F,1999Icar..140..189L}.  This model was centered on the nucleus and fitted for 120 azimuthal slices, each 3$^{\circ}$ in azimuth.  We assumed that the coma behavior is constant near the nucleus and is well-modeled by a power law.  We experimented with a range of annuli of varying positions and varying widths.  The best fit was identified by examining the remaining ``nucleus'' signal and comparing its shape to a model PSF for the WISE images using a least-squares minimization technique. We elected to use annuli fitted between nucleo-centric distances of 11$^{\prime\prime}$ and 14$^{\prime\prime}$ for the W3 image, and 13$^{\prime\prime}$ and 29$^{\prime\prime}$ for the W4 image. The model of the coma was then subtracted, leaving the nucleus signal behind.  The remaining nucleus signal is compared to the model PSF in Figure~\ref{fig:PSFcomp}.

We used aperture photometry to investigate the extracted nucleus signal. The aperture was centered on the position of the nucleus as predicted by the JPL Horizons service. The signal from the nucleus was 0.33~$\pm$~0.05~mJy and 1.56~$\pm$~0.31~mJy in W3 and W4, respectively.

The W3 and W4 signals were fit to a Near-Earth Asteroid Thermal Model (NEATM; \citealt{1998Icar..131..291H}).  The model assumes that the nucleus is spherical and acts as a smooth Lambertian surface \citep{1977Icar...31..427C, 1985Icar...64...53B}. We assumed that the emissivity of the surface was 0.9, consistent with refractory materials \citep{1986Icar...68..239L}. 
 The free-fit parameters were diameter ($D$) and beaming parameter ($\eta$).    The best fit results were $D$ = 4.135 $\pm$ 0.610 km, $\eta$ = 1.03 $\pm$ 0.21. 
 We adopted the absolute magnitude of the nucleus to be $H$ = 16.24 $\pm$ 0.02 as determined by  \cite{2006MNRAS.373.1590S} when the comet appeared to be inactive in 2005.  By coupling the absolute magnitude to the thermal fit we derive the albedo to be $p_{v}$ = 0.03 $\pm$ 0.01.  

\subsection{Coma}
\label{sec:coma}

\begin{figure}[h]
\centering
\includegraphics[totalheight=8cm]{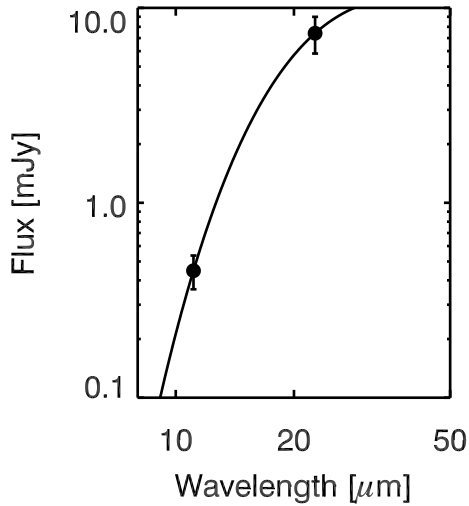}
\caption{Coma temperature fit to signal from the dust coma of 17P/Holmes.  The best-fit temperature is 134 $\pm$ 11 K, which is $\sim$~10\% higher than the local blackbody temperature.}
\label{fig:comfit}
\end{figure}

We used aperture photometry to investigate the dust coma. 
We subtracted the extracted nucleus signal reported in section~\ref{sec:nuc} from the results, giving coma fluxes of 0.45~$\pm$~0.09~mJy in W3 and 7.41~$\pm$~1.58~mJy in W4. We assumed that the signal is thermal emission, rather than reflected light, and fitted the data points at thermal wavelengths with a simple blackbody curve to determine the temperature.  The best fit is shown in Figure~\ref{fig:comfit} and corresponds to a temperature of 134 $\pm$ 11 K, $\sim$~10\% higher than the local blackbody temperature.

\subsection{Dust Trail}
\label{sec:trail}

We observed a broad dust trail lagging the nucleus of 17P in the W3 and W4 images. Given the extremely low surface brightness of the trail in W3, we restricted our analyses to the W4 data.  The trail lay in the orbital plane of 17P and was observed to stretch over 3.1~$\times$~10$^{6}$~km ($\sim$ 14$^{\prime}$.7) as projected on the sky. 

In order to determine the best model, the trail shape was first characterized using the following method: (1) Using a 40$^{\prime\prime}$ annulus centered on the comet, unwrap the image in $r$ - $\theta$ space, (2) fit a Gaussian to the unwrapped data across radial bins of 2$^{\circ}$ width, giving a location that can be converted back to x-y space, (3) repeat the process along the length of the trail. This process compresses the trail into a series of 21 discrete points that can then be analytically compared to the models described below. Several combinations of annulus width (20$^{\prime\prime}$ to 50$^{\prime\prime}$ in steps of 10$^{\prime\prime}$) and radial bin size (1$^{\circ}$, 2$^{\circ}$, and 3$^{\circ}$) were considered, with 40$^{\prime\prime}$ annuli and 2$^{\circ}$ radial bins giving the least amount of scatter in the positions.

The dust trail was modeled using the Finson-Probstein method \citep{1968ApJ...154..327F}, which assumes that the motion of cometary dust particles is controlled by Solar gravity, $F_{grav}$, and Solar radiation, $F_{rad}$. The motion can be parameterized using the ratio of the two forces, $\beta$:

\begin{equation}
\beta = \frac{F_{rad}}{F_{grav}} = \frac{5.76 \times 10^{-4}~Q_{pr}}{\rho_{d}~a_{d}}
\label{eq:betaratio}
\end{equation}

where $Q_{pr}$ is the scattering efficiency, $\rho_{d}$ is the density of the particle [kg m$^{-3}$], and $a_{r}$ is the particle radius [m]. For grains with radii larger than the wavelength of observation $Q_{pr} \sim$ 1 \citep{1979Icar...40....1B}. Thus, $\beta$ depends on the inverse of particle diameter, i.e. for smaller grains, $\beta$ is larger, meaning the radiation pressure pushing the particles outwards has a larger effect than the gravitational force pulling them inwards. We integrated the motion of the dust particles over a period of 5 years. This generated a set of points that can be shown as curves of constant radius particles released at a range of times (syndynes) or curve of constant release date with a range of particle radii (synchrones).

\begin{figure}[h]
\centering
\subfloat{\fbox{\includegraphics[totalheight=6cm]{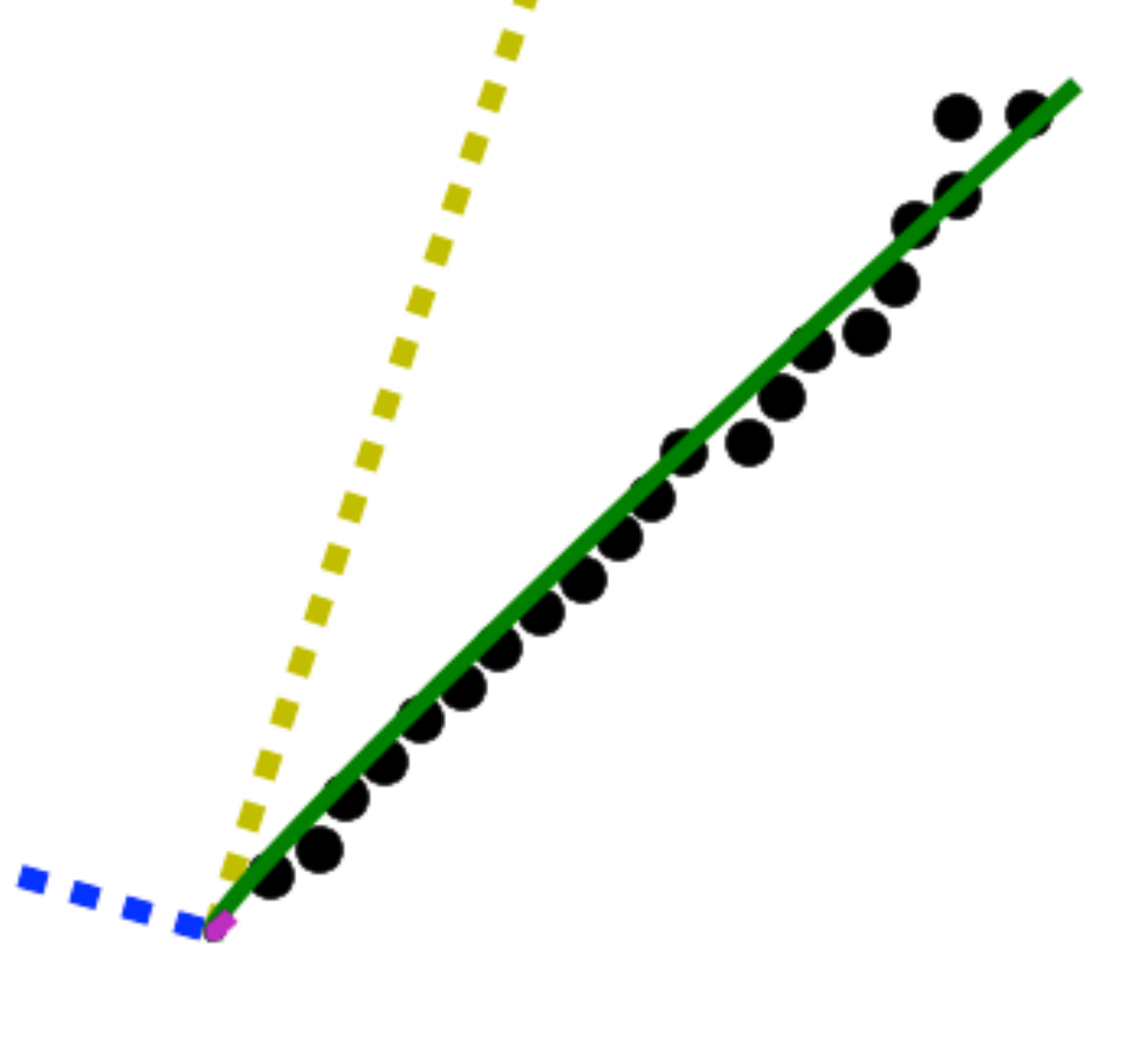}}}
\hspace{2bp}
\subfloat{\fbox{\includegraphics[totalheight=6cm]{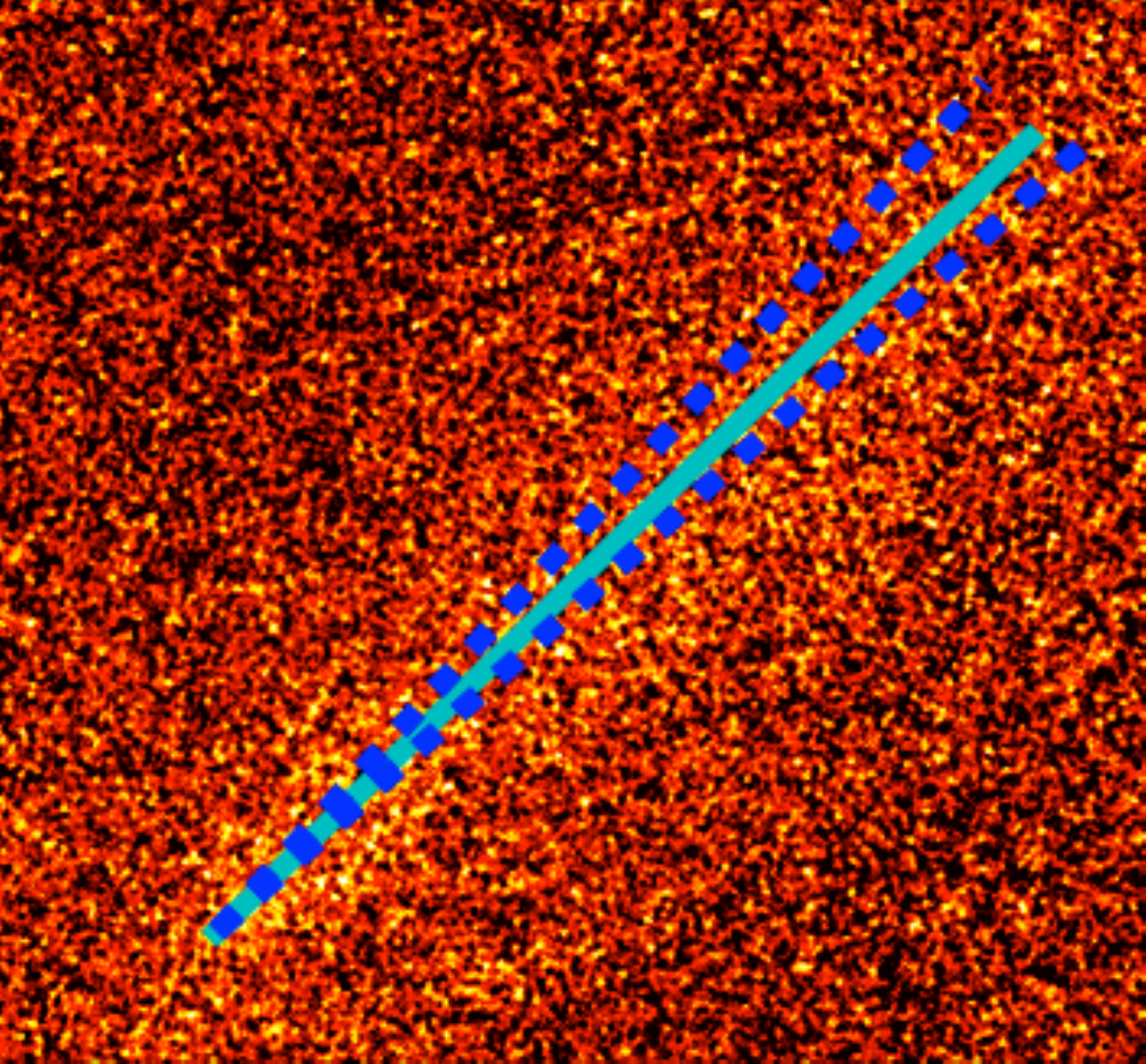}}}
\caption{Left: Schematic showing the fitted trail (black points) with two example synchrones in blue and yellow dashed lines, corresponding to particle ejection dates of 6 months and 1 year prior to observation, respectively. The solid green and magenta lines are the syndynes for particles of diameter 250~$\mu$m and 3~cm located at the apparent end of the trail and 11$^{\prime\prime}$ from the nucleus, respectively. Right: The dust trail of 17P with the best fit synchrone and error bars overplotted.  The best fit corresponds to dust that was ejected on 2007 Oct.\ 27 $\pm$ 221 days.}
\label{fig:synchrones}
\end{figure}

For both the syndynes and synchrones, we calculated the RMS between each model and the fitted tail points. The lowest RMS value for any of the syndynes is higher than for any of the synchrones, thus we proceeded to fit the trail with a synchrone. The synchrone with the lowest RMS to the fitted trail points was determined to correspond to the best-fit particle ejection date. To constrain the error on the best-fit date, we computed the best-fit synchrone for each fitted point along the trail and computed the RMS between those synchrones and the overall best-fit synchrone. This analysis yielded a best-fit ejection date of 2007 Oct.\ 27 $\pm$ 221 days. The large error bars are due to some of the fitted trail points deviating significantly from the best-fit synchrone, and thus giving dramatically different best-fit ejection dates. Example syndynes and synchrones, as well as the best fit, are plotted in Figure~\ref{fig:synchrones}. It is possible that some of the largest grains were released on previous perihelion passages as \cite{2010Icar..208..276R} observed an old debris trail along the orbit of 17P using the Spitzer space telescope in 2008. However, the contribution is likely negligible given the good fit by the synchrone analysis.

The surface brightness of the trail was too low to estimate the size distribution of particles along its length, although we can set the minimum radius of observed particles to $\sim$~250~$\mu$m.  This is done by finding the syndyne that crosses the trail near the observed edge (14.3$^{\prime}$ from the nucleus).  Using the same method, we estimate that particles near ($\sim$~11$^{\prime\prime}$ from) the nucleus are on the order of a few cm in diameter. Smaller particles likely existed beyond the trail observed by WISE but are below the detection limit.

\section{Discussion}

\subsection{Physical properties of the nucleus}

The effective nucleus diameter of 4.135 $\pm$ 0.610 km calculated here is larger than previous estimates of 3.24 $\pm$ 0.02 km \citep{2006MNRAS.373.1590S} and 3.42 $\pm$ 0.14 km \citep{2009A&A...508.1045L}, derived when 17P appeared to be inactive or only weakly active prior to its 2007 outburst. Both of the previously mentioned results were determined from optical observations using an assumed albedo of 0.04. The diameter reported here is consistent with previously reported results when they are corrected using the NEOWISE-derived albedo.  

We derived an albedo of 0.03 $\pm$ 0.01. The albedo of the nucleus is consistent with those measured for other comets, which generally occupy a narrow range between 0.02 and 0.06 \citep{2004come.book..223L}. We note that we are unable to derive an albedo for material in the trail as we do not have simultaneous  high signal-to-noise observations at optical wavelengths. \cite{2010ApJ...714.1324I} used optical, near-IR, and mid-IR observations to constrain the albedo of the ejecta within a few days of the outburst. They found that the albedo of the material (as observed at a phase angle of 16$^{\circ}$) decreased during their observations from 0.12 $\pm$ 0.04 to 0.032 $\pm$ 0.014 and suggested that sublimating volatiles would lower the albedo. \cite{2012ApJ...760L...2L} estimated the geometric albedo of the dust in the coma as 0.006 $\pm$ 0.002, also at a phase angle of 16$^{\circ}$.  Such a low value is not unheard-of \citep{2002Sci...296.1087S,2004Icar..167...37N}, though it does not match well with results from \cite{2010ApJ...714.1324I}. The discrepancy may be due to different populations of grains dominating the thermal emission in the IR and the stellar extinction at optical wavelengths. Our result for the albedo of the nucleus is generally consistent with the albedo of the ejecta observed in 2007 when the albedo was determined from combined optical and mid-IR wavelengths.

The beaming parameter of 1.03 $\pm$ 0.21 is consistent with the average value of 1.03~$\pm$~0.11 reported for 57 JFCs by the SePPCoN survey, which used thermal infrared measurements by the Spitzer Space Telescope \citep{2013Icar..226.1138F}. They found that beaming parameters for 57 JFC nuclei were approximately normally distributed and suggested that there appears to be little variation among bulk thermal properties of JFCs. As discussed in \cite{2013Icar..226.1138F}, a beaming parameter close to 1.0 implies low thermal inertia and little nightside emission. 17P appears typical in this regard.

\subsection{Thermal emission in the coma}

The best-fit coma temperature of 134~$\pm$~11~K is $\sim$~10\% warmer than the temperature expected for an ideal blackbody ($T_{BB}$) at a heliocentric distance of 5.13 AU (123~K assuming $T_{BB}$ $\propto$ 278~K~$r_{H}^{-0.5}$; \citealt{1992Icar..100..162G}). Previous results from observations taken close to the time of outburst ($r_{H}$ $\sim$ 2.4 AU) have also suggested that the dust temperature of the ejecta exceeded the local blackbody temperature of $\sim$ 180 K. \cite{2009AJ....137.4538Y} reported a dust temperature near the nucleus of 360 $\pm$ 40 K using near-IR observations obtained with the NASA Infrared Telescope Facility several days after the outburst, while mid-IR results obtained around the same time suggested cooler temperatures between 172 K and 200 K \citep{2010ApJ...714.1324I, 2009PASJ...61..679W}.  Spitzer Space Telescope observations obtained on 2007 Nov.\ 10 resulted in an estimated temperature of 260 K for the near-nucleus dust \citep{2010Icar..208..276R}. Most of these results are higher than the estimated blackbody temperature to varying degrees, matching well with our results here.

Numerous IR observations of comets have shown that it is common, perhaps even the norm, for comae and dust tail and trail temperatures to exceed the temperature expected for a co-located blackbody (e.g. \citealt{1988AJ.....96.1971T,1992Icar..100..162G,1998ApJ...496..971L,2000ApJ...538..428H}). Generally, excess emission at IR wavelengths is attributed to either small grains ($\lesssim$~1$\mu$m) that are unable to radiate efficiently at IR wavelengths, rough surfaced grains that are more emissive than the smooth spherical grains modeled by the blackbody temperature, or larger grains that maintain a thermal gradient across their surface \citep{1982come.coll..341C,1990Icar...86..236S,2001indu.book...95S}.  In the case of 17P, all of these effects may be present. The nucleus was seen to remain active in the months and years following the outburst, likely releasing small dust grains from the surface (\citealt{2012AJ....144..138S}; Snodgrass, private communication). Based on results from the best-fit synchrone determined in section~\ref{sec:trail}, particles larger than a few cm in diameter would still be close enough to the nucleus to contribute to the excess thermal emission observed here.  

\subsection{An old trail}

The morphology of the trail is consistent with being debris ejected during the 2007 outburst and observations by \cite{2013ApJ...778...19I} that showed large dust grains following the comet around aphelion in Oct.\ 2010. We constrained the range of particle diameters observed between $\sim$ 250 $\mu$m and a few cm.  The larger grain diameters are consistent with the sizes of grains observed in a slow-moving ``core'' near the nucleus just a few days after the outburst, which were determined to be $\gtrsim$~200~$\mu$m \citep{2010Icar..208..276R,2012A&A...542A..73B}. 

We measured the flux along the trail using a box aperture that extends between 11$^{\prime\prime}$ and 880$^{\prime\prime}$ from the nucleus and has a width perpendicular to the length of the trail of 52$^{\prime\prime}$. The flux was calibrated and color-corrected as described in section~\ref{sec:wobs}. To correct for light potentially lost outside of the large aperture, we applied an aperture correction of -0.03~mag derived by \cite{2013AJ....145....6J}. To estimate the cross-section of material present we used the following relation from \cite{2005Icar..179..158M}:

\begin{equation}
\sigma_{\lambda} = \frac{F_{\lambda} ~ \Delta^{2}}{B_{\lambda}(T)}
\label{eq:cross sec}
\end{equation}

where $\sigma_{\lambda}$ is the cross-section of material observed at wavelength $\lambda$ (in this case, 22~$\mu$m, or W4), F$_{\lambda}$ is the observed flux, $\Delta$ is the geocentric distance, and $B_{\lambda}(T)$ is the Planck function at temperature $T$. We were unable to constrain the temperature of the dust along the trail as the signal in W3 is too low to fit a Planck function to. We therefore assumed that the temperature is between the expected blackbody temperature of 123~K and the measured coma temperature of 134~K. This is consistent with the previously-discussed finding that many comet trails are at or exceed local blackbody temperatures. We also assumed that the temperature is constant along the trail and is not dependent on the size of the dust grains present.  The cross-section of material was 1.5~$\times$~10$^{9}$~m$^{2}$ in the case of the local blackbody temperature or 10$^{9}$~m$^{2}$ in the higher temperature case. We used previously measured minimum and maximum particle sizes ($a-$, $a+$) of 250~$\mu$m and 3~cm, and assumed that the differential size distribution of particles follows a power law of the form $n(a)~da \propto a^{-q}~da$, with the value of $q$ set between 2.2 and 3.4, as measured by \cite{2010Icar..208..276R} and  \cite{2012A&A...542A..73B}, respectively.  The mean particle size within the trail was given by:

\begin{equation}
\bar{a} = \frac{\int_{a-}^{a+} \pi a^{3} n(a) da}{\int_{a-}^{a+} \pi a^{2} n(a) da}
\label{eq:area}
\end{equation}

The mass within the observed trail was then given by:

\begin{equation}
M = \frac{4~\rho~\bar{a}~\sigma_{\lambda}}{3}
\label{eq:mass}
\end{equation}

where $\rho$ is the bulk density of the grains and was assumed to be 1000~kg~m$^{-3}$ \citep{1991ASSL..167...19J}.  The mass in the trail was $\sim$ 1.2 - 5.3 $\times$ 10$^{10}$ kg. This represented approximately 1~-~100\% of the total ejected mass \citep{2008ICQ....30....3S,2009AJ....138.1062S,2010Icar..208..276R,2010ApJ...714.1324I,2011ApJ...728...31L,2012A&A...542A..73B}. 

Assuming an average grain size of 200~$\mu$m, \citealt{2010Icar..208..276R} estimated the mass of the slow-moving core seen in 2007 to be $\sim$~4~$\times$~10$^{9}$ kg. \citealt{2012A&A...542A..73B} estimated the mass to be significantly higher at $\sim$~0.7-4~$\times$~10$^{11}$~kg by summing over an estimated particle size distribution with $a_{-}$ = 0.1~$\mu$m, 10 $< a_{+} <$ 1000 mm, and -3.3 $< q <$ -3.0. Thus, the dust trail observed by WISE represented 
3~-~75\% of the core modeled by \citealt{2012A&A...542A..73B} in 2007. 

\subsection{Why did 17P outburst?}

The overarching question remains to be answered: why does 17P undergo massive outbursts when most JFCs experience only mild mass loss? The diameter, beaming parameter, and albedo of the nucleus are similar to those of other JFC nuclei. The volatile species observed shortly after the outburst in 2007 similarly fail to provide any obvious clues about the cause of the outburst. Relative abundances of CN, C$_{2}$, C$_{3}$ and NH and the isotopic ratios of $^{12}$C/$^{13}$C and $^{14}$N/$^{15}$N in CN and HCN were similar to those observed for other comets \citep{2008ApJ...679L..49B,2009AJ....138.1062S}. Several species, including C$_{2}$H$_{6}$, HCN, CH$_{3}$OH, and C$_{2}$H$_{2}$, were enhanced with respect to H$_{2}$O although only by a factor of a few \citep{2008ApJ...680..793D}.

The perihelion distance of 17P changed from 2.17 AU in 2000 to 2.05 AU in 2007 following a close encounter with Jupiter.  The change resulted in a $\sim$~10\% increase in solar insolation at the surface. The small difference may have caused the thermal wave to propagate deeper than on previous perihelion passages, reaching previously unheated pockets of volatiles. A runaway exothermic phase transition of amorphous water ice to crystalline is probably insufficient to cause the outburst \citep{2010Icar..207..320K}. If supervolatiles such as CO and/or CO$_{2}$ are trapped within the amorphous ice and heated sufficiently, the resulting gas production may be able to drive such activity, if the gas can build up sufficient internal pressure \citep{2009AJ....138.1062S,2009ICQ....31...99S,2011Icar..212..847K,2012Icar..221..147H}.

It is possible that the nucleus of 17P has unusually high tensile strength that allows gas pressure to build up in the interior before releasing the energy in a sudden outburst upon surface failure.  \cite{2010Icar..208..276R} suggested that the nucleus must have a strength between 10~-~100~kPa in order to have survived the 2007 outburst. However, previous studies of other comets suggest much lower strengths for JFCs. 16P/Brooks 2 and D/1993 F2 (Shoemaker-Levy~9) both underwent tidal splitting during close encounters with Jupiter leading to estimates of 0.1 kPa and $\sim$ 0.38 kPa for the tensile strengths of the nuclei \citep{1985AJ.....90.2335S,1998P&SS...46...21S}. Based on observations of mini-outbursts, \cite{2008Icar..198..189B} estimated the strength of the sub-surface material of 9P/Tempel 1 to be not much more than 0.01 - 0.1 kPa, while \cite{2005Sci...310..258A} found that the strength of the surface must also be extremely low ($<$ 0.065 kPa). Only a few comets have estimated tensile strengths and are not necessarily representative of all JFCs. We note simply that the estimated tensile strength required of 17P is an order of magnitude higher than those estimated for other JFCs.

\section{Summary}

We used wide-field IR images obtained by the WISE mission in 2010 May to characterize 17P. Years later, 17P still exhibited evidence of the 2007 outburst. Our results suggest that 17P is a JFC with a typical diameter, albedo, and beaming parameter, but atypical outgassing behavior.

\begin{enumerate}

\item{The diameter, albedo, and beaming parameter of the nucleus of 17P were constrained to values of 4.135 $\pm$ 0.610 km, 0.03 $\pm$ 0.01, and 1.03 $\pm$ 0.21, respectively. The physical and bulk thermal properties of the nucleus appear to be consistent with those of other JFCs.}

\item{The temperature of dust near the nucleus was 134~$\pm$~11~K, slightly higher than the local blackbody temperature. Possible explanations for the elevated temperature include emission from small sub-micron grains that cannot effectively radiate at IR wavelengths, or contributions from larger dust grains that maintain a temperature gradient across their surface. Both effects may have been present at the time of observation.}

\item{17P was observed to have a debris trail in 2010 May. Dynamical modeling of the dust suggests that this was leftover from the massive 2007 outburst. The range of grain diameters observed is 800~$\mu$m to a few cm. The mass of trail was estimated at 1.2~-~5.3~$\times$~10$^{10}$~kg, which represents $\sim$ 1~-~100\% of the total mass ejected during the 2007 outburst.}

\end{enumerate}

\section{Acknowledgments}

This publication makes use of data products from the Wide-field Infrared Survey Explorer and NEOWISE, which is a joint project of the University of California, Los Angeles, and the Jet Propulsion Laboratory/California Institute of Technology, funded by the National Aeronautics and Space Administration. RS acknowledges support from the NASA Postdoctoral Fellowship Program.  EK was supported by the JPL Graduate Fellowship Program and the NASA Earth and Space Sciences Fellowship program.  This research was funded in part by a grant from NASA through the Near Earth Object Observations Program for the NEOWISE project, and was carried out at the Jet Propulsion Laboratory, California Institute of Technology, under a contract with the National Aeronautics and Space Administration.

\end{document}